# Silicene Nanomesh

Feng Pan[1,4,†], Yangyang Wang[3,†], Kaili Jiang[3,†], Zeyuan Ni[3], Jianhua Ma[5], Jiaxin Zheng[6], Ruge Quhe[3,7], Junjie Shi,[3] Jinbo Yang,[2,3] Changle Chen[1], and Jing Lu[2,3,*]

[1]Shaanxi Key Laboratory of Condensed Matter Structures and Properties, School of Science, Northwestern Polytechnical University, Xi'an 710072, P. R. China

[2]Collaborative Innovation Center of Quantum Matter, Beijing 100871, P. R. China

[3]State Key Laboratory for Mesoscopic Physics and Department of Physics, Peking University, Beijing 100871, P. R. China

[4]School of Physics and Telecommunication Engineering, Shaanxi University of Technology, Hanzhong 723001, P. R. China

[5]School of Physics and Nuclear Energy Engineering, Beihang University, Beijing 100191, P. R. China

[6]School of Advanced Materials, Peking University, Shenzhen Graduate School, Shenzhen 518055, P. R. China

[7]Academy for Advanced Interdisciplinary Studies, Peking University, Beijing 100871, P. R. China

[†]These authors contributed equally to this work.

[*]Address correspondence to jinglu@pku.edu.cn

**Abstract**

Similar to graphene, zero band gap limits the application of silicene in nanoelectronics despite of its high carrier mobility. By using first-principles calculations, we reveal that a band gap is opened in silicene nanomesh (SNM) when the width $W$ of the wall between the neighboring holes is even. The size of the band gap increases with the reduced $W$ and has a simple relation with the ratio of the removed Si atom and the total Si atom numbers of silicene. Quantum transport simulation reveals that the sub-10 nm single-gated SNM field effect transistors show excellent performance at zero temperature but such a performance is greatly degraded at room temperature.

Silicene, silicon analog of graphene, is predicted to possess a Dirac-cone-shaped energy band[1] and ultra-high carrier mobility[2], and thus has a potential application in high-performance nanoelectronics. Recently silicene has been successfully grown on Ag[3-7], $ZrB_2$[8], Ir[9], and $MoS_2$[10] substrates. However, zero band gap of pristine silicene limits its application as a logic element in electronic devices directly. It is critical to open the band gap of silicene without degrading its carrier mobility. Several methods have been proposed to open the band gap of silicene from first-principles calculations, such as vertical electric field[2,11,12], surface adsorption[13,14], or semihydrogenation[15]. Nonetheless, an experimentally approachable vertical electric field can only open a gap below 0.1 eV in silicene, evidently smaller than the minimum band gap requirement (0.4 eV) for traditional field effect transistors (FETs)[2]. Metal atom adsorption is able to induce a larger band gap up to 0.66 eV in silicene[13,14,16], but a large supply voltage ($V_{dd}$) of about 1.7 ~ 30 V is required [13,14]. It is highly desirable to design a new silicene FET with a high on/off ratio under a low supply voltage.

Computing technology requires a channel length of FET smaller than 10 nm in next decades. However, bulk-Si FET will not perform at sub-10 nm channel length because of its short-channel effects[17]. In order to enable continued FET scaling, one can modify Si device structure or use an alternative channel geometry/material. Up to now, 8 ~ 10 nm advanced Si FETs (including Si nanowire with gate-all-round-configuration[18], double-gated FinFET[19], and extremely thin Si on insulator (ETSOI)[20]) and 9 nm carbon nanotube (CNT) FETs[21] have been fabricated. It is interesting to examine whether silicene FETs are competitive with these existing advanced Si FETs and CNTFETs.

In this Article, we propose a novel method to open the band gap of silicene, namely, fabrication of silicene nanomesh (SNM). Theoretical calculations[22-32] have shown that the band gap of graphene can be opened by making a periodic array of holes (namely nanomesh), and the size of the gap depends on the structural parameters of graphene nanomesh (GNM). Experimentally, a transport gap has been observed in GNM FETs with a triangular array of hexagonal holes[33-37], and the on/off current ratio can reach 100 [38], which is one order of magnitude larger than that of pristine graphene FETs. Based on density functional theory (DFT) calculations, a band gap is opened up in SNM, whose size strongly depends on the width $W$ of the wall between the neighboring holes and has a maximum value of 0.68 eV.

Subsequently, we simulate the transport properties of the sub-10 nm SNM FETs based on the non-equilibrium Green's function (NEGF). The simulated SNM FETs show excellent device performance with an on/off ration up to $10^4$ at a supply voltage of 0.5 V. When phonon scattering is considered, the performance is greatly degraded with an on/off ration down to 100.

## Results and Discussion

### Geometry and electronic structure

The SNM model is built by digging a simple triangular array of hexagonal holes in a silicene sheet, as shown in Fig. 1(a). The edge of the holes has zigzag shape and the edge Si atoms of the holes are passivated by hydrogen atoms. Each type of SNM is designated by the notation [$R$, $W$], where the $R$ index reflects the radius of the hole calculated by $N_{removed} = 6R^2$ ($N_{removed}$ is the number of the removed Si atoms from one lattice cell) and the $W$ index is the width of the wall between the nearest-neighboring holes. Fig. 1(a) shows an example of [$R$, $W$] = [1, 4] SNM.

After relaxation, the unit cell size of the SNM structures is nearly unchangeable compared with the corresponding silicene supercell. However, constructing periodic holes will slightly affect the buckling distance ($\Delta$) of the edge silicon atoms, which is larger than that of pristine silicene (0.46 Å) and increases with the increase of $W$ given the same $R = 1$ (Fig. S1(a)). For the silicon atoms far away from the holes, the $\Delta$ tends to be the value of pristine silicene, as shown in Fig. S1(b).

To investigate the stability of SNMs, the cohesive energy $E_{coh}$ and Gibbs free energy $\delta G$, are calculated. The per-atom cohesive energy $E_{coh}$ is calculated according to the follow equation:

$$E_{coh} = (N_{Si}E(Si) + N_H E(H) - E(SNM))/(N_{Si} + N_H) \qquad (1)$$

where $E$(Si) and $E$(H) are the energies of the free silicon and hydrogen atoms, respectively, $E$(SNM) the total energy of a SNM in one supercell, $N_{Si}$ and $N_H$ the numbers of silicon and hydrogen atoms in a SNM supercell, respectively. The calculated $E_{coh}$ for [$R = 1$, $W$] SNM is positive (3.64 ~ 4.71 eV) and increases with the increasing $W$ as shown in Fig. 2a. We define the per-atom Gibbs free energy ($\delta G$) of formation for SNMs as

$$\delta G = -E_{\text{coh}} + n_{\text{Si}}\mu_{\text{Si}} + n_{\text{H}}\mu_{\text{H}} \qquad (2)$$

where $n_{\text{Si}}$ and $n_{\text{H}}$ are the mole fraction of Si and H atoms, respectively, for a given structure, and $\mu_{\text{Si}}$ and $\mu_{\text{H}}$ are the per-atom chemical potentials of Si and H, respectively, at a given state. We chose $\mu_{\text{Si}}$ and $\mu_{\text{H}}$ as the binding energies per atom of bulk Si and H$_2$ molecule, respectively. As given in Fig. 2b, the calculated $\delta G$ values for $W = 1\sim3$ are 0.012~0.075 eV, and therefore a small amount of additional energy is required to make these reaction processes happen. However, SNMs have a negative $\delta G$ when $W > 3$, indicating a higher thermodynamical stability relative to their elemental reservoirs. Based on the width dependence of $E_{\text{coh}}$ and $\delta G$, the stability of SNMs monotonically increases with the increasing $W$ given the same $R = 1$. To further study its stability, a molecule dynamic simulation of the $[R = 1, W = 2]$ SNM is performed at temperature of 1000 K. As shown in the Supplementary Movie S1 we added, the structure is well kept, suggesting SNMs are stable enough against the high temperature.

The calculations done by Liu et al.[26] show that the band gap of graphene hexagonal nanomeshes is only opened when $W$ is even. While $W$ is odd, the GNMs behave semimetallically like pristine graphene. Our calculations show that SNMs have similar properties. The band structures of $[1, W]$ SNMs with $W = 1 \sim 10$ are provided in Fig. 1(b) and 1(c). Similar to pristine silicene, the SNMs with odd $W$ show semimetallic behavior, with a pair of linear bands crossing at the $K$ point (Fig. 1(b)). By contrast, a direct band gap ($E_g$) is opened at the $\Gamma$ point when $W$ is even (Fig. 1(c)). The band structures of $[2, W]$ SNMs with even $W$ are provided in Fig. S2 and the band gaps are also opened in them. The opened band gap when $W$ is even monotonically decreases with the increasing $W$ given the same $R$. A maximum band gap of about 0.68 eV is observed in both $R = 1$ and 2 cases (Fig. 1(d)).

Pedersen et al.[25] found that the band gap $E_g$ in GNM is determined by the relation

$$E_g = g \frac{N_{\text{removed}}^{1/2}}{N_{\text{total}}}, \qquad (3)$$

where $N_{\text{total}}$ and $N_{\text{removed}}$ are the numbers of the total Si atoms before digging the holes and the removed hole atoms in a unit cell, respectively, and $g$ is a fitting factor. For GNM, one has $g = 25$ eV. Fig. 1(e) shows the band gap of SNM against $N^{1/2}_{\text{removed}}/N_{\text{total}}$. The linear relation remains with $g = 7.246$ eV, which is much smaller than that for GNM. Therefore, given the same $N_{total}$ and $N_{removed}$, *i.e.* with the same notation [R, W], the band gap in SNM is much smaller than that in GNM. The band gap opening in GNM is proved not directly caused by quantum confinement as in graphene nanoribbions (GNRs); instead, it has a geometric symmetry origin[32,37]. When the two reciprocal lattice vectors of a GNM overlap with Dirac points of the pristine graphene, degeneracy at the Dirac points is lifted and a sizable band gap appears; otherwise, it's semimetal like graphene. SNM shares similar mechanism of band gap opening with GNM. In [R, W] SNMs, when W is even its two reciprocal lattice vectors overlap with Dirac points of the pristine silicene, *i.e.* the K and K' points of pristine silicene are folded into the Γ points of SNM. Due to the intervalley scattering, a band gap is opened in SNMs when W is even.

$E_g \propto \frac{1}{N_{total}}$ can be explained by if we approximately treat SNM as a periodic potential perturbation $U(r)$ to pristine silicene. If $K - K' = G$, where $G$ is one reciprocal lattice vector of the SNM supercell, there is interaction between the two degenerate Dirac points. As a result, a band gap is opened and is expressed in terms of degeneracy perturbation theory as,

$$E_g = 2|V_G| = \frac{2}{S}\int_{supercell}\psi_A^K(r)^*U(r)\psi_A^{K'}(r)dr = \frac{2}{S}\int_{supercell}U(r)e^{-i(K-K')\cdot r}\mu_A^K(r)^*\mu_A^{K'}(r)dr \quad (4)$$

where $S$ is the square of the supercell of SNM, $\psi_A^{K*}(r)$ and $\psi_A^{K'}(r)$ are the Bloch function of A sublattice at the K and K' points, with periodic part of $\mu_A^K(r)$ and $\mu_A^{K'}(r)$, respectively. The external periodical potential $U(r)$ induced by the hole is assumed to be rather localized. Given the same hole of different SNMs, $\int_{supercell}U(r)e^{-i(K-K')\cdot r}\mu_A^K(r)^*\mu_A^{K'}(r)dr$ is approximately independent of the size of the supercell, and we therefore have $E_g \propto \frac{1}{N_{total}}$ due to $N_{total} \propto S$.

Fig. 3 presents the effective mass ($m^*$) of the conduction band bottom of SNMs along the Γ→K ($m_e^{\Gamma K}$) and Γ→M ($m_e^{\Gamma M}$) directions as a function of W. The effective mass is calculated by using the formula:

$$m^* = \eta^2\left(\frac{\partial^2 E(k)}{\partial k^2}\right)^{-1}. \quad (5)$$

The effective mass $m^*$ monotonically decreases from 0.093 to 0.022 $m_0$ for $R = 1$ and from 0.151 to 0.034 $m_0$ for $R = 2$ with the increasing W because of the reduced band gap, where $m_0$ is the free electron mass. $m_e^{\Gamma K}$ is approximately equal to $m_e^{\Gamma M}$ except for $W = 2$ case. At the same W, the $m^*$ values with $R = 2$ are slightly larger than their respective $m^*$ values with $R = 1$.

**Transport properties of SNM FETs**

The schematic model of a single-gated FET based on the [1, 2] SNM is presented in Fig. 4(a). The electrodes are composed of semi-infinite silicene. To avoid the interaction between SNM and $SiO_2$ dielectric, a $h$BN buffer layer is inserted between the SNM and $SiO_2$ substrate,[2] and the thickness of $SiO_2$ dielectric plus $h$BN buffer region is $d_i = 11$ Å. To start with, we calculated the transmission spectrum of a 6.5 nm-gate-length SNM FET by using the DFT method with single-$\zeta$ (SZ) basis set to benchmark our SE extended Hückel result (Fig. S3). The transmission spectra calculated between the two methods are similar, except that the size of the transmission gap generated by the SE method (0.9 eV) is a litter larger than that by the DFT method (0.7 eV) and the transmission coefficients generated by the SE method near the Fermi level are unsmooth and generally larger than those by the DFT method. The larger transmission gap generated by the SE method can cause the decrease of on/off ratio compared with that by the DFT method given the same gate voltage window. The on/off ratio may be further slightly decreased when using the SE method due to the relative larger conductance in the off-state contributed by the larger transmission coefficient near $E_f$. However these won't affect much the results. Then we focus on the transport properties of the SNM FET with a larger gate length $L_{gate} = 9.1$ nm. The conductance in SNMs can be modulated by applying a gate voltage to the channel, and an on/off switch is expected.

The transmission spectra of the 9.1 nm-gate-length SNM FET at $V_g = 0$, and 0.5 V with $V_{bias} = 0.2$ V are presented in Fig. 4(b). When $V_g = 0$ V, there is a transport gap of 0.9 eV centered at the Fermi level ($E_f$). The transmission coefficient nearly vanishes within the bias window, indicating an off state. By applying a positive gate voltage, the transport gap can be shifted towards low energy direction. At $V_g = 0.5$ V, relatively large transmission coefficients are moved inside the bias window. According to Eq. (1), the drain current is calculated and then normalized by the channel width to obtain the current density $I_{ds}$ (Fig. 4(c)). Clear on/off current modulation is achieved. If we set $V_{dd} = V_{on} - V_{off} = 0.5$ V and $V_g = 0.5$ V is chosen as the on-state, the on/off ratio can reach $5.1 \times 10^4$, which is about three orders of magnitude larger than the maximum on/off ratios obtained in dual-gated silicene FET[2] and already meets the requirement of $10^4 \sim 10^7$ for the high-speed logic applications. The subthreshold swing (SS, here is defined as $dV_{gate}/d(\log I)$) is 68 mV/dec, which approaches the 60 mV/dec switching limit of the classical transistors. To provide an insight into the switch capability, we investigate the transmission eigenchannels of the off-state ($V_g$ = 0 V) and on-state ($V_g$ = 0.5 V) at $E$ = 0.05 eV and $k$ = (0, 0), as shown in Fig. 4(d). The transmission eigenvalue of the off-state is merely 6.71 × $10^{-7}$, and the corresponding incoming wave function is obviously scattered and unable to reach to the other lead. On the contrary, the transmission eigenvalue of the on-state is 0.78; as a result, the scattering is weak, and the most of the incoming wave is able to reach to the other lead.

To determine the scaling effect of the gate length $L_{gate}$ on the device performance, we calculate the transfer characteristics of the SNM FET with different gate lengths (3.8 ~ 9.1 nm) at a fixed bias voltage of $V_{bias}$ = 0.2 V as shown in Fig. 5. The maximum current $I_{max}$ is insensitive to $L_{gate}$. By contrast, the minimum current $I_{min}$ increases with the decreasing $L_{gate}$. Such a scaling behavior is attributed to the increasing off-state leakage current with the decreased $L_{gate}$. Therefore, the maximum and minimum current ratio $I_{max}/I_{min}$ decreases significantly from 5.8 × $10^5$ at $L_{gate}$ = 9.1 nm to 1.9 × $10^2$ at $L_{gate}$ = 3.8 nm (Fig. 6(a)). The on/off current ratio $I_{on}/I_{off}$ (the gate voltage window is limited to a supply voltage) is a more important parameter than $I_{max}/I_{min}$ to characterize switching effect of an electronic device. We limit the gate voltage window to 0.5 V and show the $L_{gate}$ dependent $I_{on}/I_{off}$ in Fig. 6(b). It also monotonously decreases from 5.1 × $10^4$ at $L_{gate}$ = 9.1 nm to 17 at $L_{gate}$ = 3.8 nm.

The subthreshold swing SS = $dV_{gate}/d(\log I)$ is another important parameter of FET and determines how effectively the transistor can be turned off by changing the gate voltage. The SS of the SNM FETs monotonously increases from to 68 to 336 mV/dec when $L_{gate}$ scales down from 9.1 to 3.8 nm. Transconductance $g_m$ is another important parameter to characterize switching effect of an electronic device, which is computed from $g_m = \partial I_{ds}/\partial V_g$. The $g_m$ value decreases from 555 μS/μm at $L_{gate}$ = 3.8 nm to 351 μS/μm at $L_{gate}$ = 9.1 nm (Fig. 6(d)). Another key parameter the intrinsic gate capacitance $C_g$ is calculated in Fig. 6(e). $C_g$ is defined as $C_g = \partial Q_{ch}/\partial V_g$, where $Q_{ch}$ is the total charge of the channel. The relationship between the $C_g$ and $L_{gate}$ is the following equation[39]: $C_g = \varepsilon_0 \varepsilon_r W_{gate} L_{gate}/t_{ox}$, where $\varepsilon_0$ and $\varepsilon_r$ are the dielectric constant of vacuum and the relative dielectric constant of the gate dielectric, $W_{gate}$ is the width of the gate, and $t_{ox}$ is the thickness of the gate dielectric. As shown in Fig.

6(e), $C_g$ indeed increases almost linearly with $L_{gate}$ from 159 aF/μm at $L_{gate}$ = 3.8 nm to 258 aF/μm at $L_{gate}$ = 9.1 nm. According to the charge control model at low bias, $\mu = \frac{g_m}{V_{bias}} \cdot \frac{L_{gate}^2}{C_g}$, since $C_g \propto L_{gate}$, if we assume the mobility is a constant, the transconductance $g_m$ tends to vary inversely with $L_{gate}$, and our results is consistent with this tendency.

Fig. 6(f) shows $L_{gate}$ dependence of charge carrier transit time $\tau$ based on the calculated $C_g$ and $g_m$, i.e. $\tau = C_g/g_m$. $\tau$ increases from 0.29 to 0.73 ps when $L_{gate}$ increases from 3.8 to 9.1 nm. The intrinsic cut-off frequency $f_T$ indicates how fast the channel current is modulated by the gate and is described as $f_T = 1/(2\pi\tau)$[40-42]. $f_T$ decreases monotonically with $L_{gate}$ from 557 GHz at $L_{gate}$ = 3.8 nm to 217 GHz at $L_{gate}$ = 9.1 nm (Fig. 6(g)), which is much smaller compare with that in the sub-10 nm graphene FETs (4 ~ 22 THZ)[43]. The drift velocity of a transistor can be derived by $v_{drift} = L_{gate}/\tau$. As shown in Fig. 6(h), $v_{drift}$ is insensitive to the gate length and is 12.4 × 10$^5$ ~ 13.3 × 10$^5$ cm/s when $L_{gate}$ = 3.8 ~ 9.1 nm.

Future FET technologies will require operation at voltages at or below 0.5 V to reduce power consumption. To compare the SNM FETs with the Si based and CNT transistors at a supply voltage $V_{dd}$ = 0.5 V, we summarize the critical performance parameters of the sub-10 nm SNM (9.1 and 7.8 nm), advance Si, and CNT FETs at $V_{bias}$ = 0.5 V in Table 1 . The 9.1 nm SNM carries an on-state current of 464 μA/μm, which is larger than those (41 ~ 300 μA/μm) of the 8 ~ 10 nm advanced Si devices but slightly smaller than that (630 μA/μm) of the 9 nm CNT device. The on/off current ratio of the 9.1 nm SNM FET is 7.4 × 10$^3$, which is a little smaller than those (1 × 10$^4$) of the 10 nm Si nanowire, 8 nm ETSOI, and 9 nm CNT devices but larger than that (1 × 10$^3$) of 10 nm Si Fin device. The SS value (82 mV/dec) of the 9.1 nm SNM FET is slightly smaller than those (83 ~ 125 mV/dec) of the 10 nm Si nanowire and Si Fin, 8 nm ETSOI, and 9 nm CNT devices. Taking the three criterions together, the 9.1 nm SNM FET is competitive with the sub-10 nm advanced Si devices but is inferior to the 9 nm CNT device.

Adding the total area of the gates is an effective way to strengthen the gates' control over the channel, the gate control ability of a FET is expected to be improved by using a dual gate configuration. The transfer characteristic of the 9.1 nm dual-gated SNM FET at $V_{bias}$ = 0.5 V is provided in Fig. 7(a) to compare with that of the single-gated one with the same $L_{gate}$, and improved gate control is apparent. The performance parameters of the 9.1 nm dual-gated SNM FET are generally better than those of the single-gated counterpart as listed in Table 1. The SS is reduce by 8 meV/dec, the large on/off current ratio is increased by a factor of 2.7, and the on-state current is increased by a factor of 1.6. The 9.1 nm dual-gated SNM delivers an on-state current of 870 $\mu A/\mu m$, which is larger than those (41 ~ 630 $\mu A/\mu m$) of the 8 ~ 10 nm advanced Si devices and 9 nm CNT device. The on/off current ratio of the 9.1 nm dual-gated SNM is $1.2 \times 10^4$, which is comparable with those of the 10 nm Si nanowire, 8 nm ETSOI, and 9 nm CNT devices and one order of magnitude larger than that of the 10 nm Si Fin device. The SS (74 mV/dec) of the 9.1 nm dual-gated SNM is smaller than those (83 ~ 125 mV/dec) of the 8 ~ 10 nm advanced Si devices and 9 nm CNT device. Taking the three criteria together, the 9.1 nm dual-gated SNM has a better performance than the sub-10 nm advanced Si devices and 9 nm CNT device. The excellent performance of the SNM FET is attributed to the depressed short channel effects due to their extremely small thickness and fewer traps on semiconductor-dielectric interface due to the smooth interface (Fig. 8).

The output characteristics for the 9.1 nm SNM FET at different gate voltages are shown in Fig. 7(b). The source-drain ballistic current increase with the applied bias voltage, and no current saturation is observed until $V_{bias} = 0.7$ V. The current of the dual-gated SNM FET is much larger than that of the single-gated SNM FET at the same $V_{bias}$ under $V_g = 0.5$ V, indicating an improved gate controlling.

The transfer and output characteristics of the 7.8 nm SNM FET is provided in Fig. S4. Although the performance of the 7.8 nm single-gated SNM FET is inferior to the sub-10 nm advanced Si devices and the 9 nm CNT device, the 7.8 nm dual-gated SNM FET is sufficiently improved: The SS is reduce by 4 meV/dec to 68 mV/dec, the large on/off current ratio is increased by a factor of 2.7 to $8.9 \times 10^3$, and the on-state current is increased by a factor of 3 to 607 $\mu A/\mu m$ at a supply voltage of 0.5 V (Table 1). Consequently, the 7.8 nm dual-gated SNM has a better performance than the sub-10 nm advanced Si devices and is competitive with the 9 nm CNT device.

It is interesting to examine whether the SNM FETs can meet the requirements for the high-performance FETs from the 2013 edition of the International Technology Roadmap for Semiconductors (ITRS)[44]. The required gate lengths of HP logic of 2022 and 2023 are 8.9 nm and 8.0 nm, and supply voltages is 0.72 and 0.71 V, respectively. The transfer characteristics of the 9.1 nm dual-gated SNM FET at $V_{bias}$ = 0.72 V is provided in Fig. 6(a). The 9.1 nm dual-gated SNM FET, whose gate length is approximately meet the requirement of HP logic of 2022 (8.9 nm), carries an on-state current of 3122 $\mu$A/$\mu$m at a supply voltage of $V_{dd}$ = 0.72 V and greatly satisfies the requirement of $I_{on}$ = 1350 $\mu$A/$\mu$m for the HP logic of ITRS of 2022. As shown in Fig. S4(a), the calculated on-state current (1963 $\mu$A/$\mu$m) of the 7.8 nm dual-gated SNM FET at a supply voltage of $V_{dd}$ = 0.71 V also meets the requirement ($I_{on}$ = 1330 $\mu$A/$\mu$m) of the HP logic of ITRS of 2023. Unfortunately, the on/off current ratios of the 9.1 and 7.8 nm dual-gated SNM FETs are only $1.8 \times 10^3$ and $1.2 \times 10^3$, respectively. Both of them cannot meet the requirement of HP logic of ITRS ($1.33 \times 10^4$ in 2022 and $1.35 \times 10^4$ in 2022).

We perform a molecular dynamics (MD) simulation of the channel region in a 9.1 nm single-gated SNM FET at room temperature to check how the transport properties change as phonon effect is partially included (only elastic scattering is considered) in the device. Compared with the transmission spectra of 9.1 nm single-gated SNM FET at $V_{bias}$ = 0.2 V without considering the phonon scattering, the transport gap is increased from 0.9 eV to 1.5 eV, and the transmission coefficients of both the conduction and valence bands are greatly depressed at 300 K after phonon scattering effect is included, as shown in Fig. 9. The off-state current at $V_g$ = 0 V isn't affected much. Whereas the on-state current is decreased significantly to $3.9 \times 10^{-2}$ $\mu$A/$\mu$m, and the on/off current ratio is decreased to 39 at $V_{bias}$ = 0.2 V when the gate bias window is fixed at 0.5 V. When $V_{bias}$ = 0.5 V, the on-state current is decreased to 4.0 $\mu$A/$\mu$m, and the on/off current ratio is decreased to 100. Therefore, SNM FETs still works at room temperature, but its performance is greatly degraded. Phonon scattering plays an import role on accurate assessment of SNM FETs even at a short-gate length below 10 nm.

In summary, a band gap is opened in SNM when the width $W$ of the wall between the neighboring holes is even from the first-principles calculations. The size of the band gap increases with the reduced $W$ and is proportional to the ratio of the removed Si atom and the total Si atom numbers of silicene. We simulate the transport of the FETs with a sub-10 nm SNM channel based on quantum transport theory and find that the sub-10 nm SNM FETs have an excellent performance at zero temperature, characterized by a large on-state current up to 870 $\mu$A/$\mu$m, a large on/off current ratio up to $1.2 \times 10^4$, and a small subthreshold swing low to 68 mV/dec at a supply voltage of 0.5 V. However the performance is greatly degraded when phonon scattering effect is included.

**Methods**

The geometry optimizations and the band structure calculations are performed using the double numerical basis set plus polarization (DNP), implemented in the DMol$^3$ package[45]. We chose the generalized gradient approximation (GGA)[46] of the Perdew-Burke-Ernzerhof (PBE) form to the exchange-correlation functional[47]. Both the atomic positions and lattice constant are relaxed without any symmetry constraints until the maximum force is smaller than 0.01 eV/Å. A 16 × 16 × 1 Monkhorst-Pack $k$-points grid[48] is used in the first Brillouin zone sampling. A vacuum space of 20 Å normal to silicene plane is used to avoid spurious interaction between periodic images. To examine the thermal stability of SNMs, *ab initio* MD simulation within the NVT ensemble is carried out using the DMol$^3$ package at 1000 K, and the process lasts for more than 1.0 ps with a time step of 1.0 fs.

A single-gated two-probe model is built to simulate the transport of SNM, and the pristine silicene is used as source and drain electrodes for simplicity. Transport properties are calculated by the semi-empirical (SE) extended Hückel method coupled with NEGF formalism implemented in the Atomistix Tool Kit (ATK) 11.2 package[49-51]. The Hoffman basis is used, and the temperature is set at 300 K. The $k$-point meshes of the electrodes and central region are set to 1 × 50 × 50 and 1 × 50 × 1, respectively. The current is calculated with the Landauer-Büttiker formula[52]:

$$I(V_g, V_{\text{bias}}) = \frac{2e}{h} \int_{-\infty}^{\infty} \{T(E, V_g, V_{\text{bias}})[f_L(E - \mu_L) - f_R(E - \mu_R)]\} dE \qquad (6)$$

where $T(E, V_g, V_{\text{bias}})$ is the transmission probability at a given gate voltage $V_g$ and bias voltage $V_{\text{bias}}$, $f_{L/R}$ the Fermi-Dirac distribution function for the left(L)/right(R) electrode, and $\mu_L/\mu_R$ the electrochemical potential of the L/R electrode.

To include the phonon effect in the calculation of transport properties, *ab initio* MD simulation of the central region of the device within the NVT ensemble is performed by using the Dmol[3] package at 300 K, and the process lasts for 3.0 ps with the electrode extension parts constrained. The time step is 1.5 fs. Then different configurations of the central region are built into two-probe models after every 400 MD steps, and their transport properties are evaluated and finally averaged over 5 configurations using a NEGF approach implemented in the ATK package.

**Acknowledgements**

This work was supported by the National Natural Science Foundation of China (Nos. 11274016, 51072007 and 61471301), the National Basic Research Program of China (Nos. 2013CB932604 and 2012CB619304), National Foundation for Fostering Talents of Basic Science (Nos. J1030310/ J1103205), Program for New Century Excellent Talents in University of MOE of China, and the Special Fund of Education Department of Shaanxi Province, China (Grant No. 2013JK0635).

**Author Contributions**

J.L. conceived the idea. F.P. and Y.W. did the transport simulations. K.J. performed the electronic calculations. The data analyses were performed by F.P., Y.W., K.J., Z.N. and J.M.. J.Z., R.Q., J.S., J.Y. and C.C. helped discussing. This manuscript was written by F.P., Y.W. and J.L. All authors reviewed this manuscript.

**Additional information**

**Supplementary information** accompanies this paper at http://www.nature.com/Scientificreports

**Competing financial interests:** The authors declare no competing financial interests.

**Corresponding author:** Jing Lu, jinglu@pku.edu.cn

**Table 1**. Comparison of Performance Metrics Between Sub-10 nm SNM, Advance Si, and CNT Transistors.

| Channel | $L_{ch}$ (nm) | $V_{bias}$ (V) | $V_{dd}=V_{on}-V_{off}$ (V) | $I_{on}$ (μA/μm) | $I_{on}/I_{off}$ | SS (mV/dec) |
|---|---|---|---|---|---|---|
| SNM (single-gated) | 7.8 | 0.5 | 0.5 | 205 | $3.3\times10^3$ | 72 |
| SNM (dual-gated) | 7.8 | 0.5 | 0.5 | 607 | $8.9\times10^3$ | 68 |
|  |  | 0.71 | 0.71 | 1963 | $1.2\times10^3$ | 74 |
| SNM (single-gated) | 9.1 | 0.5 | 0.5 | 464 | $7.4\times10^3$ | 82 |
| SNM (dual-gated) | 9.1 | 0.5 | 0.5 | 870 | $1.2\times10^4$ | 74 |
|  |  | 0.72 | 0.72 | 3122 | $1.8\times10^3$ | 113 |
| Si nanowire[30] | 10 | 0.5 | 0.5 | 300 | $1.0\times10^4$ | 89 ($V_{bias}$ = 1.0V) |
| Si Fin[19] | 10 | 0.5 | 0.5 | 138 | $1.0\times10^3$ | 125 ($V_{bias}$ = 1.2 V) |
| ETSOI[20] | 8.0 | 0.5 | 0.5 | 41 | $1.0\times10^4$ | 83 ($V_{bias}$ = 1.2 V) |
| CNT[21] | 9.0 | 0.5 | 0.5 | 630 | $1.0\times10^4$ | 94 |
| HP logic[a] | 8.9 | 0.72 | 0.72 | 1350 | $1.35\times10^4$ |  |
| HP logic[b] | 8.0 | 0.71 | 0.71 | 1330 | $1.33\times10^4$ |  |

HP logic technology requirements of ITRS in [a]2022 and [b]2023.

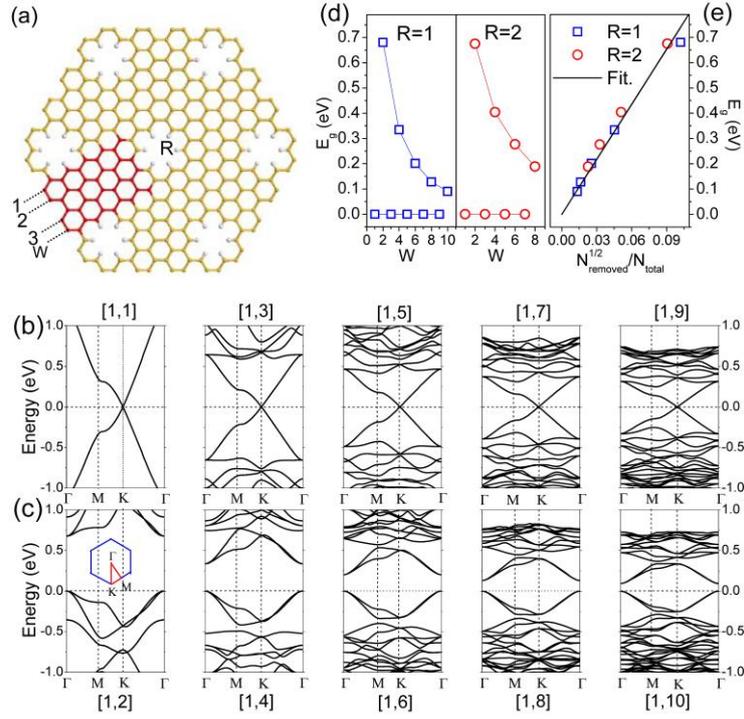

**Figure 1**. (a) Geometric configuration of SNMs with $R = 1$ and $W = 4$. The yellow and white balls stand for Si and H atoms, respectively. Energy band structures of SNMs with odd (b) and even (c) $W$. (d) The band gap of SNM as a function of $W$ with $R = 1$ and 2. (e) Band gap is plotted versus the quantity $N_{removed}^{1/2}/N_{total}$ ($N_{removed}$ is the number of removed Si atoms, and $N_{total}$ is equal to the number of atoms before the holes is made in the unit cell). The black line represents a linear fit to the data.

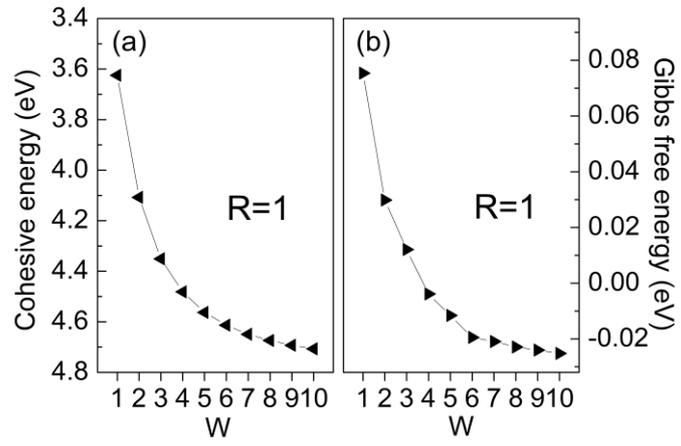

**Figure 2**. Per-atom cohesive energy $E_{coh}$ (a) and Gibbs free energy $\delta G$ (b) in SNMs as a function of $W$ with $R = 1$.

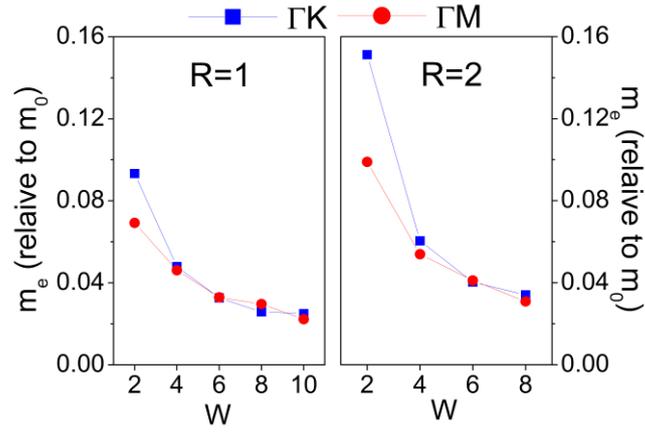

**Figure 3**. Effective mass of electron in SNMs at the conduction band bottom along the Γ→*K* and Γ→*M* directions as a function of *W* with *R* = 1 and 2.

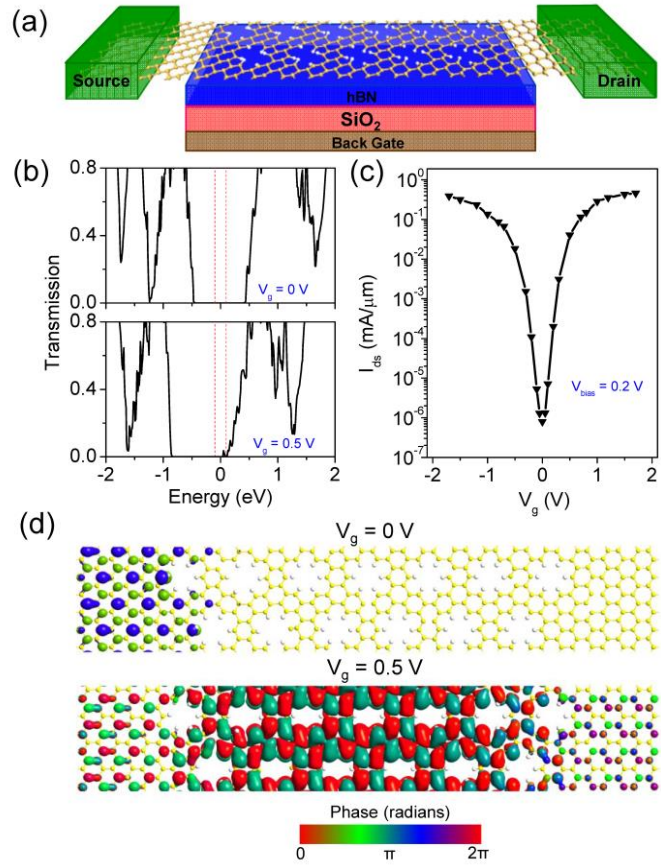

**Figure 4**. (a) Schematic of a graphdiyne device structure. The yellow and white balls stand for Si and H atoms, respectively. (b) Transmission spectra of the SNM transistor with $L_{gate}$ = 9.1 nm at $V_g$ = 0 and 0.5 V. The bias voltage is fixed at $V_{bias}$ = 0.2 V. The vertical red dashed-lines denote the bias voltage window. The Fermi level is set to zero. (c) Transfer characteristic of the device at $V_{bias}$ = 0.2 V. (d) Transmission eigenstates of the off-state ($V_g$ = 0 V) and on-state ($V_g$ = 0.5 V) at $E$ = 0.05 eV and $k$ = (0, 0). The isovalue is 0.2 a.u.

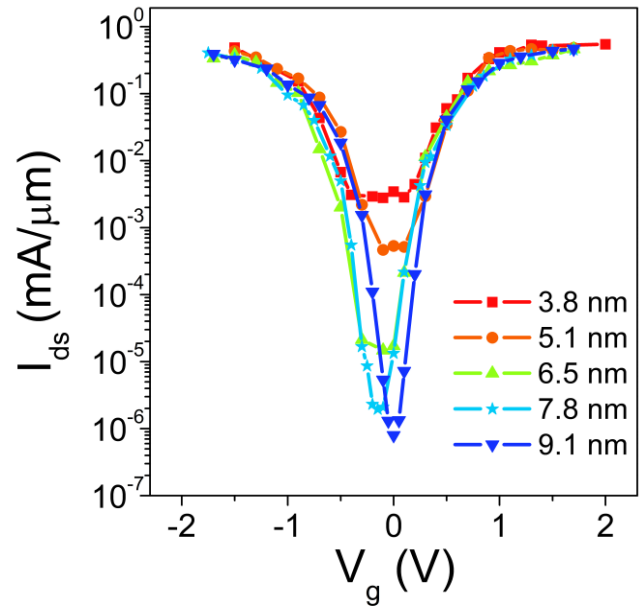

**Figure 5**. Transfer characteristics of the SNM FETs for different channel lengths at $V_{bias} = 0.2$ V.

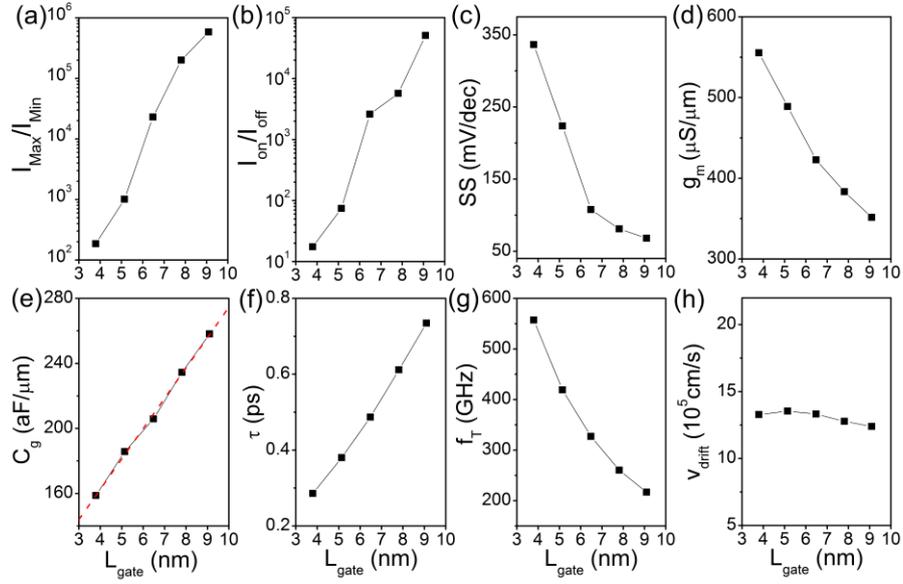

**Figure 6.** Critical device performance parameters of the SNM field effect transistors as a function of the gate length at $V_{bias} = 0.2$ V: (a) $I_{max}/I_{min}$ current ratio, (b) on/off current ratio at a gate voltage window of 0.5 V, (c) subthreshold swing, (d) transconductance, (e) intrinsic gate capacitance, (f) charge carrier transit time, (g) intrinsic cut-off frequency, and (h) carrier drift velocity.

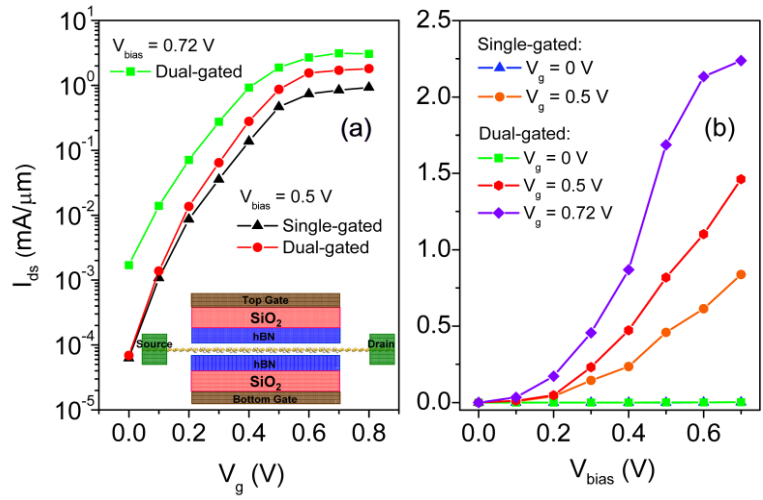

**Figure 7.** (a) Transfer characteristics of the 9.1 nm dual-gated SNM FET at $V_{bias}$ = 0.5 and 0.71 V compared with that of the single-gated SNM FETs with the same $L_{gate}$ at $V_{bias}$ = 0.5 V. The inset is the side view of the dual-gated SNM transistor device. (b) Output characteristics for the 9.1 nm single- and dual-gated SNM FETs at different gate voltages.

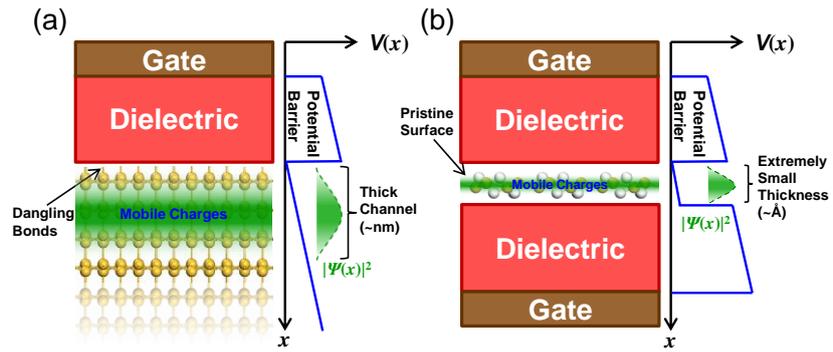

**Figure 8.** Schematic of FET channel made of traditional 3D material (Si) and 2D material (SNM) and their corresponding vertical potential diagrams. 2D materials have clean surface with fewer traps in semiconductor-dielectric interface and are extremely thin compared to traditional 3D materials, leading to a better gate control. $V(x)$ and $|\Psi(x)|^2$ represent the potential and the probability density of the electronic charges, respectively.

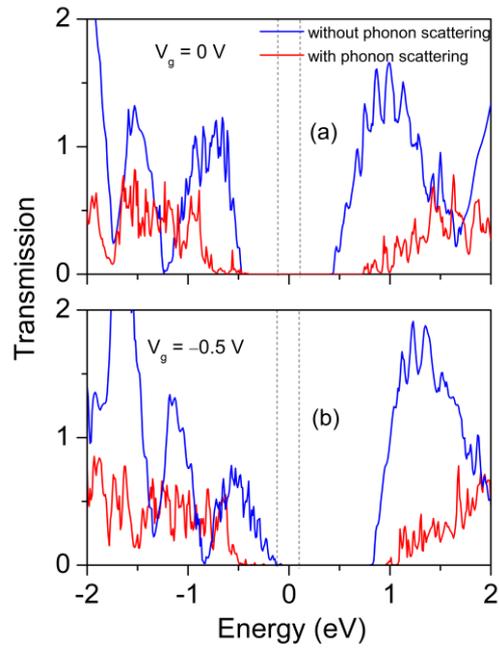

**Figure 9.** Transmission spectra of the 9.1 nm single-gated SNM FET at $V_g = 0$ and $-0.5$ V with (averaged over five different times) and without considering the phonon scattering under $V_{bias} = 0.2$ V. The vertical black dashed-lines denote the bias voltage window. The Fermi level is set to zero.